\documentclass[10pt]{article}
\usepackage{amsmath}
\usepackage{amssymb}

\usepackage{graphicx}

\usepackage{cite}

\usepackage{color} 

\usepackage[labelfont=bf,labelsep=period,justification=raggedright]{caption}

\makeatletter
\renewcommand{\@biblabel}[1]{\quad#1.}
\makeatother

\date{}

\begin{document}

% Title must be 150 characters or less
\begin{flushleft}
{\Large
\textbf{Algorithmic choice of coordinates for injections into the brain:
encoding a neuroanatomical atlas on a grid}
}
% Insert Author names, affiliations and corresponding author email.
\\
Pascal Grange, 
Partha P. Mitra 
\\
\bf Cold Spring Harbor Laboratory,\\
 One Bungtown Road, Cold Spring Harbor, New York 11724, USA
\\
$\ast$ E-mail: pascal.grange@polytechnique.org
\end{flushleft}

% Please keep the abstract between 250 and 300 words
\section*{Abstract}
 
Given an atlas of the brain and a number of injections to be performed in
order to map out the connections between parts of the brain, we propose an algorithm
to compute the coordinates of the injections. The algorithm is designed to sample 
the brain in the most homogeneous way compatible with the separation of brain regions. 
 It can be applied to other species for which a neuroanatomical atlas is available.    
The computation is tested on the annotation at a resolution of 25 microns
corresponding to the Allen Reference Atlas, which is hierarchical 
and consists of 209 regions. The resulting injections are being used for 
the injection protocol of the Mouse Brain Architecture project. Due to its large size
and layered structure, the cerebral cortex is treated in a separate algorithm, which is more adapted to
its geometry.\\

\section*{Introduction}

Neuroanatomy is experiencing a renaissance thanks to the 
 huge amount of data generated through molecular biology 
and image-processing \cite{AllenAtlasMol, images}. Brain regions can be
separated in a data-driven way and compared to brain regions defined 
by classical neuroanatomy \cite{clusters, markerGenes}. Ongoing data generation at the Mouse Brain Architecture project
aims to construct the full matrix of directed connections between parts of the brain 
\cite{MBA}. In order to do so, various tracers are injected {\emph{in vivo}} into the brain
of a mouse. Cryosectioning of the brain after a survival period produces 
a series of images showing the projections of the injection sites.\\

The present note presents the algorithm used to compute the
coordinates of the injections.  The problem we address is a coding problem in the sense
that a three-dimensional object, with complex internal structure, has
to be represented by a set of points, and the number of points in the set is much smaller 
than the number of voxels in the numerical atlas encoding the internal structure. The algorithm is not
specific to the atlas or number of injections we considered. The injection
 coordinates can be recomputed if a different number of injections is chosen,
or if a different atlas is used. This allows the protocol to be
extended to other species for which a neuroanatomical atlas is
available.  For species that do not have a cerebral cortex for instance, the
special treatment of the cortex that is made in one of the loops of
the algorithms will simply not be executed. The corresponding
    Matlab code is available upon request, to be executed using an atlas
    formatted into a three-dimensional grid by the user.\\

Given an atlas of the  brain, we have to compute the coordinates of
$I$ injections in the left hemisphere\footnote{As the left hemisphere is better-charted in the neuroanatomy
  literature, it was decided to perform injections in that hemisphere
  only, but the algorithm can be used for the whole brain, or for any
  region of the brain for which an atlas is available.}  in such a way
that these injection sites encode the atlas.  The computation of
coordinates has to satisfy the following two
constraints:\\ 1. {\bf{Sampling constraint.}} In order to explore the
brain efficiently, injections must not be too far from each
other.\\ 2. {\bf{Separation constraint.}} In order to assign the
origin of the tracers to a definite brain region, injections must not
be too close to the boundaries of the regions.\\

\begin{figure}
\begin{minipage}{0.43\textwidth} 
The results of the algorithm in the left hemisphere of the mouse brain, with special treatment 
of the cerebral cortex, can be observed on a coronal section (Figure (\ref{fig:NaccAndCortex}))
containing cortical injections, placed as columns of five 
pulses intersecting the average surface of the cortex, at sites
 sampling the average surface of the cortex. This coronal section also contains two injections 
targeting nucleus accumbens.
From this figure it is clear that the local grids of injections
going into each regions are slightly displaced in the rostrocaudal direction wrt
the grid of injections into nucleus accumbens. Otherwise, all the other targeted regions, apart from cerebral cortex 
and nucleus accumbens, would have injection sites in this coronal plane. The boundary of nucleus accumbens therefore introduces an irregularity  into the grid of injections.\\
A 3D plot of the injection targets, shown on  Figure (\ref{fig:dislocations}), looks locally like a crystal with many irregularities induces by boundaries of brain regions. On top of this crystal, the cortical injections look like a bunch of ordered spines.\\
\end{minipage}
\begin{minipage}{0.49\textwidth} 
\centering
\includegraphics[width=3.5in]{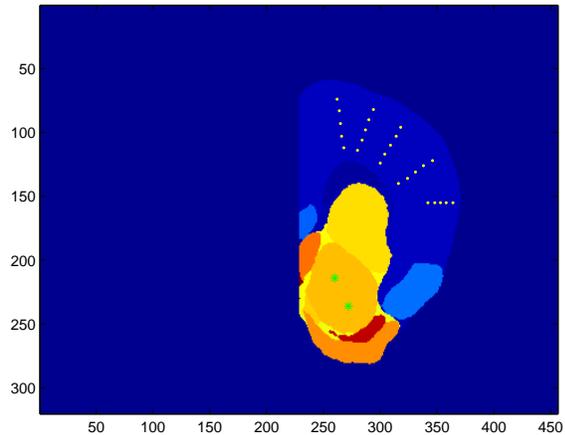}
\begin{center}
\caption{A coronal section with five 'columns of injections' in the cerebral cortex (shown in yellow)  and
two injections into nucleus accumbens (shown in green).}
\label{fig:NaccAndCortex}
\end{center}
\end{minipage}
\end{figure}

\begin{figure}
\centering
\includegraphics[width=4in]{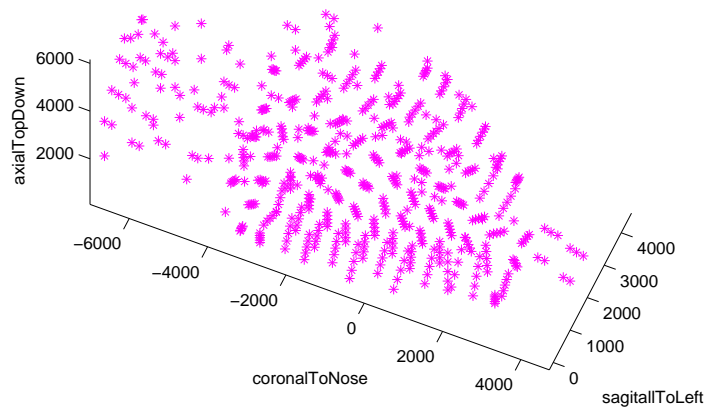}
\caption{The injection targets shown as a cloud of magenta stars.}
\label{fig:dislocations}
%\end{minipage}
\end{figure}

 If the brain had no
internal structure, the atlas would consist of a single big
region. Both constraints would be satisfied by filling the brain with
a hexagonal sphere packing, with some exclusion zone around the
surface of the brain. Injections would be placed at the centers of
the spheres. They would form a crystal structure. The algorithm we
propose applies this method to regions within the brain. The whole set
of injections is therefore not quite regular: it resembles a crystal
with defects induced by the boundaries of brain regions. Across a
large brain region, the injections sites will look like a crystal. In
a cluster of very small brain regions, they will look like an
amorphous pile of points, with the coordinates of the points dictated
mostly by the geometry of the boundaries of the regions (see Figures (\ref{fig:dislocations}) and 
(\ref{fig:NaccAndCortex}) for illustrations).\\

The structure of the paper is as follows. The algorithm used to choose regions 
and to compute coordinates is first described in pseudocode. Relevant orders of magnitudes
for the sizes of brain regions in the Allen Reference Atlas are then derived. This leads
to an estimate of a reasonable security distance to be used to satisfy the separation constraint.
 The computational techniques used to estimate distances from boundaries are then exposed. They are
 based on solutions to the eikonal equation. The cerebral cortex is treated in a separate 
way due to its large size and layered structure. Results are presented for the mouse brain.
 We used the Allen Reference Atlas\cite{AllenAtlas},
  which consists of 209 regions, with a resolution of 25 microns, intersected with the left hemisphere of the brain\footnote{As the
 left hemisphere is better-charted in the neuroanatomy literature, it was 
decided to perform injections in that hemisphere only, but the algorithm can be used for the whole brain or for any other regions
of the brain for which an atlas is available.}.
 The value $I = 250$ was
  dictated by time constraints on the production of data for the Mouse
  Brain Architecture project. The algorithm can be used to encode 
  other atlases, with another number $I$ of injection sites. Geometric properties of the resulting injection 
coordinates (pairwise distances and distances to boundaries of brain regions)
are then studied in order to check that the sampling and separation constraints are resonably 
satisfied.

% Results and Discussion can be combined.
\section*{Algorithm}

Given an atlas $\mathcal{A}$ of the left hemisphere of the brain (i.e. a hierarchical partition of the left 
hemisphere of the brain, endowed with a natural tree structure), and given 
the total number of injections $I$ to be placed, 
 the algorithm reads as the following pseudocode:\\

1. Choose the targeted brain regions.\\
1.1 Compute the critical size for a targeted region: $V_{inj}=V_{tot}/I$, where
$V_{tot}$ is the total volume of the left hemisphere of the brain.\\
1.2 For each region in the atlas $\mathcal{A}$ that is a leaf of the tree, compute its volume. 
If it is larger than $V_{inj}$, declare 
the region to be one of the targeted regions, and cut the corresponding leaf.\\
Otherwise lump the region with is parent region on the tree, and cut the corresponding leaf.\\
1.3 Repeat the procedure until the root of the tree is reached.\\

2. Define a security distance $\sigma$: distances between injections and boundaries\footnote{Distances to surfaces 
are measured using the eikonal distance, described below in more details.}
of neuroanatomical regions should be larger than $\sigma$ in order to satisfy the separation constraint. The
value of $\sigma$ is estimated below for the left hemisphere of the mouse brain.\\  

3. For each targeted region $r$:\\
3.1 Compute the number $I_r$ of injections to be placed in the region, under the constraint
$\sum_r I_r = I$.\\
3.2 If $I_r = 1$, place the injection at the centroid of the region.\\ 
    Otherwise, intersect the region with a sphere packing, and adjust 
    the radius of the spheres so that exactly $I_r$ centers of spheres
    are in the region, and at distances from the boundary that exceed $\sigma$.
    If this fails, retry with a reduced value of sigma.\\

Given the large size of the cerebral cortex and its layered structure,
it deserves a separate study in the mouse brain. Injections into the
cortex are performed as a series of pulses along the trajectory of the
needle. Given the sampling and separation constraints, these trajectories must be distributed as 
line segments intersecting an average layer of the cerebral cortex at
sites that are homogeneously distributed on the layer. This gives rise to a intrinsically
two-dimensional algorithm, described in a separate subsection, to be
executed in the case of cerebral cortex only. It should be skipped
when the algorithm is applied to species for which the delineation of
the cerebral cortex is problematic or impossible.

\subsection*{Choice of targeted regions}

%In order to generate the tex files for the tables to be included here, run
%write_tables_of_annotation( Ref, AnnotationTwoSided )

We used a numerical version of the  Allen Reference Atlas on a grid
of size $528\times320\times456$, with voxel side 25 micrometers. The volume 
of a voxel is therefore roughly 0.016 nanoliter.
At this resolution, 209 regions of the left hemisphere can be detected. For each of
them, we computed the percentage of the volume of the left hemisphere it occupies.\\

As $I$ injection sites have to
  be chosen in the atlas ($I=250$ for the Mouse Brain Architecture project),
 a region is declared to be too small to be targeted 
  if its volume is smaller than the typical injection volume defined
  as $V_{inj}=V_{tot}/I$, where $V_{tot}$  is the volume of the brain (or of the 
left hemisphere of the brain if the atlas is restricted to the left hemisphere).

\subsection*{Security distances from boundaries of regions}
In order to separate the regions, the injection sites should not be
too close to the boundary of any of the regions. In order to ensure
this in a systematic way, we need to define a security distance,
denoted by $\sigma$. We would like to forbid the placement of any
injection site within $\sigma$ of the boundary of any region (however,
this can fail for a fixed value of $\sigma$ in the case of regions that are very thin in one
direction; the algorithm handles such cases by decreasing sigma
until the sites have some room to be included). We present a simple
scaling argument to derive an estimate of $\sigma$ that serves as an
initial value of the security distance in the algorithm. But first we
have to define the way distances to surfaces are computed.\\

Let us denote by $\mathcal{V}$ the space occupied by the region of
interest, say caudoputamen, in the left hemisphere. Its boundary is a surface, denoted by 
$\mathcal{B}$. As we need to know how far the injection sites are from the boundary, we need to 
compute the distance $h$ between every point of $\mathcal{V}$ and the boundary. This is achieved by
solving the eikonal equation in $\mathcal{V}$, with boundary conditions on $\mathcal{B}$:
\begin{equation}
|\nabla h | = 1\;\; {\mathrm{in}}\; {\mathcal{V}}, 
\label{eq:eikonal}
\end{equation}
$$ h = 0\;{\mathrm{on}}\; {\mathcal{B}}.$$
The eikonal equation is solved using level-set methods, such as the ones described in \cite{Sethian}.
If we think of the surface as an object emitting light,
the value of the solution $h$ at a given point is the shortest optical path followed by light coming
from the object, provided the light propagates in a medium with index 1 (the index of the medium is actually the 
r.h.s. of the eikonal equation, as this equation is the Euler--Lagrange equation associated to an action equal 
to the optical path \cite{LLMec}). Another intuitive interpretation of the eikonal equation (\ref{eq:eikonal}) is in terms of combustion: if the boudary $\mathcal{B}$
is set on fire, and the fire propagates in a homogeneous medium, $h( x )$ is the time at which point $x$ in the medium 
will be touched by the fire. This interpretation gives rise to the level-set algorithm in a natural way.\\ 

Having solved the eikonal equation, we can assess how well the
separation constraint is satisfied.  Injections should be comfortably
far from the boundary of the region. So we define a {\emph{security
    region}} given by points inside $\mathcal{V}$ whose distance from
the boundary is larger than a {\emph{security value}}, denoted by
$\sigma$, which can be estimated based on the packing density of a regular hexagonal
sphere packing. As the packing density for a hexagonal
packing is
$$d = \frac{\pi}{3\sqrt{ 2}},$$
a first guess for the security value corresponds to a situation where spheres of radius $\sigma$ are hexagonally 
packed in the volume $V_{tot}$, without any dislocation 
$$ I \times \frac{4 \pi}{3} \sigma^3 = d \times V_{tot},$$
i.e.
$$\sigma = \left( \frac{V_{tot}}{ 4\sqrt{2} I }\right)^{1/3} \simeq
500\;{\mathrm{microns}}$$ for the left hemisphere of the mouse brain
with $I=250$, which can be difficult to achieve for very thin
regions. Hence this value will be the initial value of the security
distance in the algorithm, and will be lowered until the desired
number of injections can be hexagonally packed in the region, with all
the distances to boundary larger than the security value. In
particular, for the region labelled 'Striatum', which consists of
several small subregions of striatum lumped together (larger
subregions such as nucleus accumbens and caudoputamen, are targeted
regions and receive several injections of their own), the security
distance has to be reduced.\\

For each targeted region, given the solution $h$ of the eikonal
equation with boundary condition at the boundary of the region
(meaning internal boundary on a grid), the set of voxels within the
region for which $h$ is larger than $\sigma$ will be referret to as
the {\emph{security region}}.  As some of the hemisphere will not be
used for injections because it does not fall into any security region,
the minimal values of distances to boundaries obtained at the end of
the algorithm will be smaller on average than the initial value of
$\sigma$. See the discussion section for statistics on the distances
to boundaries.\\

\subsection*{Computation of injection sites in a given brain region}

In every brain region (apart from the cerebral cortex that will be treated
separately, the injections are placed at the centers of speres in a
hexagonal packing of regular spheres, whose radius is calibrated so
that there are the desired number of injection sites in the region.
 There is some freedom in translations, dilations and rotations of the 
packing, which has not been used so far to maximize the average distance to the boundary, as the first 
configuration found was accepted.\\

If there is just one injection to be made in the region, its coordinates are taken to be those of the 
 centroid $C_r$ of the 
 security region ${\mathcal{S}}_r$, which is the point  within the security region that minimizes the variance:
$$C_r = {\mathrm{argmin}}_C\sum_{x\in{\mathcal{S}}_r}( x - C)^2.$$
In case the security region ${\mathcal{S}}_r$ is convex, the coordinates of $C_r$ are just the 
averages of the coordinates of the points in ${\mathcal{S}}_r$, as the average is the quantity that minimizes the variance.\\

If more injections are to be placed in the region, a precomputed grid
of injections is superposed onto the security regions (the axes of the
packing coincide with those of the grid containing the atlas). The
intersection between the grid and the security region contains a
number of points that depends on just one parameter, which is the
radius of the sphere packing. The algorithm tries a range of values of
this parameter until the intersection contains exactly the desired
number of injections.  It can be that the number of points in the
intersection jumps by large quantities, thereby missing the desired
number of injections.  This occurs for very singular, anisotropic
regions. In such cases, the security distance $\sigma$ is lowered
until a suitable value for the radius of the sphere packing exists.\\

\subsection*{Results for the mouse brain, using the Allen Reference Atlas and $I=250$ injections}

We used a numerical version of the  Allen Reference Atlas on a grid
of size $528\times320\times456$, with voxel side 25 micrometers. The volume 
of a voxel is therefore roughly 0.016 nanoliter.
At this resolution, 209 regions of the left hemisphere can be detected. For each of
them, we computed the percentage of the volume of the left hemisphere it occupies.
The results are contained in the tables at the end of the present note.\\

As $I=250$ injection sites have to
  be chosen in the atlas, a region is declared to be too small
  if its volume is smaller than the typical injection volume defined
  as $V_{inj}=V_{tot}/I$, where $V_{tot}\simeq 0.19\;{\mathrm{mL}}$  is the volume of the left hemisphere of the mouse
  brain. The value of $V_{inj}$ is roughly $0.7$ microliters, corresponding
to 0.4 percent of the volume of the left hemisphere.\\

As described in step 1.2 of the algorithm, one starts with the leaves
of the hierarchical annotation with 209 indices. If the volume of a
leaf is smaller than the critical volume $V_{inj}$, it is lumped with
its parent, otherwise the leaf is declared to be one of the targeted
regions. The leaves are then 'cut' and this procedure is repeated
until the root of the tree is reached.  Some regions in the Allen
Reference ATlas \cite{AllenAtlas}, such as 'Basic Cell Groups and
Regions', 'Cerebrum' and 'Brain Stem', do not form a clear anatomical pattern.
 We discarded them, and as their volumes add up to eight times
the volume $V_{inj}$ taken by a single injection, eight injections were
added to the number of injections into the cerebral cortex.\\

The first table of brain regions, Figure (\ref{fig:standardLevel7}), shows a list of the leaves big enough to be targeted.
The first column contains the name of the brain region, the second column, called 'Nb inj.' contains the 
number of injections to be placed in the region, which is the closest integer to the quotient 
of the volume of the region by the critical volume $V_{inj}$. The third column, called 'Index', contains the index of the region, 
which can be used to access the region in the full table of regions in the annotation.
The fourth column, called 'Parent', is the index in the main table of the parent of the region. The last column,
called 'LumpedIndices', gives the list of indices of regions (drawn from the full table), that have been lumped with the region because they are smaller than $V_{inj}$. As we are at the level of leaves in the annotation, the values contained in the third and fifth columns are the same.\\

\begin{figure}
\centering
\noindent
\begin{tabular}{|l|l|l|l|l|}
\hline
\textbf{Name}&\textbf{Nb Inj.}&\textbf{Index}&\textbf{Parent}&\textbf{LumpedIndices}\\\hline
Main olfactory bulb&10&6&5&6\\\hline
Anterior olfactory nucleus&2&8&5&8\\\hline
Piriform area&7&10&5&10\\\hline
Ammon's Horn&8&17&16&17\\\hline
Dentate gyrus&3&18&16&18\\\hline
Subiculum&2&20&19&20\\\hline
Caudoputamen &13&24&23&24\\\hline
Nucleus accumbens &3&26&25&26\\\hline
Olfactory tubercle &2&28&25&28\\\hline
Lateral septal nucleus &1&30&29&30\\\hline
Pallidum- dorsal region&1&37&36&37\\\hline
Substantia innominata&2&40&38&40\\\hline
Ventral posterior complex of the thalamus&1&82&80&82\\\hline
Zona incerta&1&110&107&110\\\hline
Inferior colliculus&3&113&112&113\\\hline
Superior colliculus- sensory related&1&118&112&118\\\hline
Superior colliculus- motor related&3&120&119&120\\\hline
Cerebellar cortex&26&209&204&209\\\hline
\end{tabular}

\caption{{\bf{Results of the first loop of the algorithmic choice of targeted regions.}} Leaves of the 'standard' annotation (at 'level 7' in the tree) whose volumes are larger than
$V_{inj}$, with the numbers of injections they receive.}
\label{fig:standardLevel7}
\end{figure}

\begin{figure}
\centering
\includegraphics[width=4in]{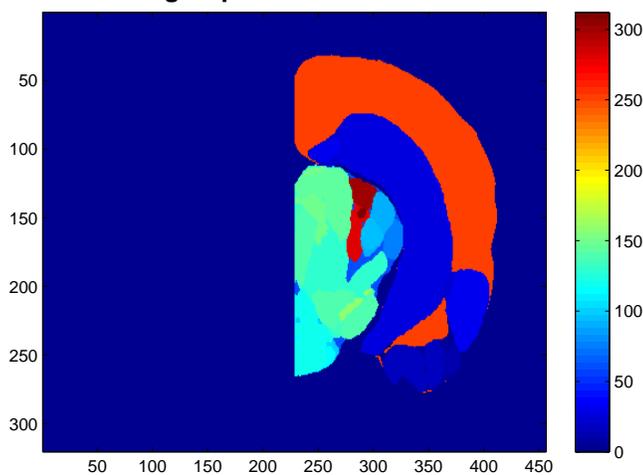}
\caption{A coronal section of the standard annotation with non-empty intersection with the lateral group of the dorsal thalamus 
(value 259 in the color bar), the lateral posterior nucleus of the thalamus (value 279 in the color bar), 
and the suprageniculate nucleus (value 309 in the color bar). These three brain regions, when lumped together,
make up a region just big enough to be targeted by one injection.}
\label{fig:lumped}
\end{figure}

The result of the next step of the algorithm is presented in another
table, Figure (\ref{fig:standardLevel6}), organized in the same
way. For instance one can see that the lateral group of the dorsal
thalamus (index 57), was lumped with the lateral posterior nucleus of
the thalamus (index 58) and the suprageniculate nucleus (index
59). The volumes of those three regions add up to $\simeq 0.48$
percentage points of the left hemisphere, which is just above the the
fraction represented by $V_{inj}$, thus leading to one injection for
this group of structures. Indeed the second column shows one
injection. A coronal section intersecting the three regions is shown
on figure (\ref{fig:lumped}). As the annotation is hierarchical, the
resulting targeted region is connected.\\
\begin{figure}
\centering
\noindent
\small{
\begin{tabular}{|l|l|l|l|l|}

\hline
\textbf{Name}&\textbf{Nb Inj.}&\textbf{Index}&\textbf{Parent}&\textbf{LumpedIndices}\\\hline
Olfactory areas&4&5&4&5   7   9  11  12  13  14\\\hline
Retrohippocampal region&7&19&15&19\\\hline
Striatum-like amygdalar nuclei&2&32&22&32  33  34  35\\\hline
Anterior group of the dorsal thalamus&1&49&48&49  50  51  52  53  54  55\\\hline
Lateral group of the dorsal thalamus&1&57&48&57  58  59\\\hline
Ventral group of the dorsal thalamus&1&80&75&80  81\\\hline
Periaqueductal gray&2&124&119&124  125\\\hline
Pons- sensory related&2&149&148&149  150  151  152  153\\\hline
Pons- motor related&2&154&148&154  155  156  157  158  159  160  161  162\\\hline
Vestibular nuclei&1&195&184&195  196  197\\\hline

\end{tabular}
}

\caption{{\bf{Results of the second loop of the algorithmic choice of targeted regions.}}  Leaves of the 'standard' annotation (at 'level 6' in the tree) whose volumes are larger than
$V_{inj}$ after being lumped with the leaves at level 7 that are too
small to be targeted, with the numbers of injections they receive. }
\label{fig:standardLevel6}
\end{figure}
\begin{figure}
\centering
\noindent
\begin{tabular}{|l|l|l|}
\hline
Brain Region & Volumetric fraction & Nb injections\\
\hline
Cerebral cortex & 0.31 & 66 \\
Retrohippocampal region & 0.04 & 10 \\
Hippocampal reg. &  0.04 & 10\\
Olfactory areas &  0.10 & 25 \\
Medulla & 0.06 & 15 \\
Cerebellum &  0.12 & 30 \\
Thalamus &  0.05 & 13\\
Pons &  0.05 & 13\\
Striatum &  0.09 & 23 \\
Hypothalamus & 0.04 & 10 \\
Midbrain & 0.08 & 20 \\
Pallidum &  0.02 & 5 \\
\hline 
\end{tabular} 
\caption{The twelve big regions in the left hemisphere (coarsest non-hierarchical partition), 
with fractions of the volume occupied by each region, and numbers of injections received.}
\label{fig:big12Broken}
\end{figure}
Full details of the whole hierachical procedure are not included in the present note,
however the hierarchical atlas is compatible with the (non-hierarchical) partition of the left hemisphere 
into 12 regions, also provided by the Allen Reference Atlas \cite{AllenAtlas}. The number of injections
 into each of  those twelve regions, as shown in a separate table, Figure (\ref{fig:big12Broken}), is rougly proportional to the fraction of the volume of the left hemisphere it occupies. 
\begin{figure}
\centering
\includegraphics[width=4in]{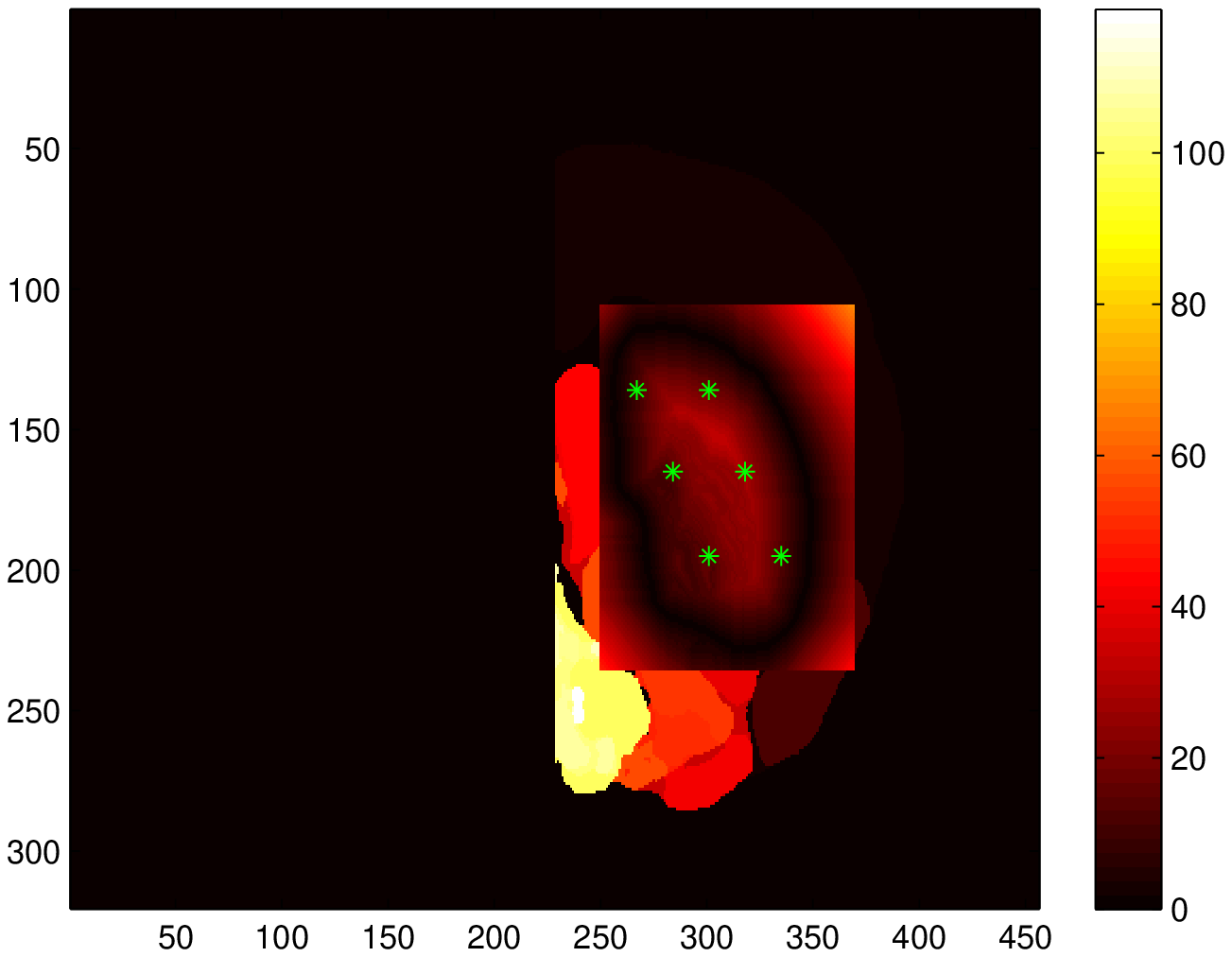}
\caption{{\bf{A coronal section of the caudoputamen, showing distances to the boundary and injection sites.}} A coronal section of the eikonal distances to the surface of the caudoputamen (computed in a small
region enclosing the caudoputamen), overlaid 
onto the atlas, with six injections displayed as green stars. Distances in the 
colorbar are expressed in multiples of the voxel side, which is 25 microns. The characteristic level-set
 structure of the solution to the eikonal equation is manifest. The boundary of the caudoputamen in this section is at level 0. 
}
\end{figure}
Since the targeted regions are all treated separately by the algorithm,
the injection grid looks locally like a crystal, but irregularities are induced
by the boundaries between targeted brain regions. The locally hexagonal structure of the packing is
visible in a coronal section of caudoputamen containing six of the thirteen injection
sites computed in this structure. The eikonal distance function to the boundary 
of caudoputamen was only computed in a box containing caudoputamen, for the sake of speed.
 As the algorithm uses only values of the eikonal function inside the brain region.
\subsection*{The special case of the cerebral cortex}
\begin{figure}
\centering
\includegraphics[width=4in]{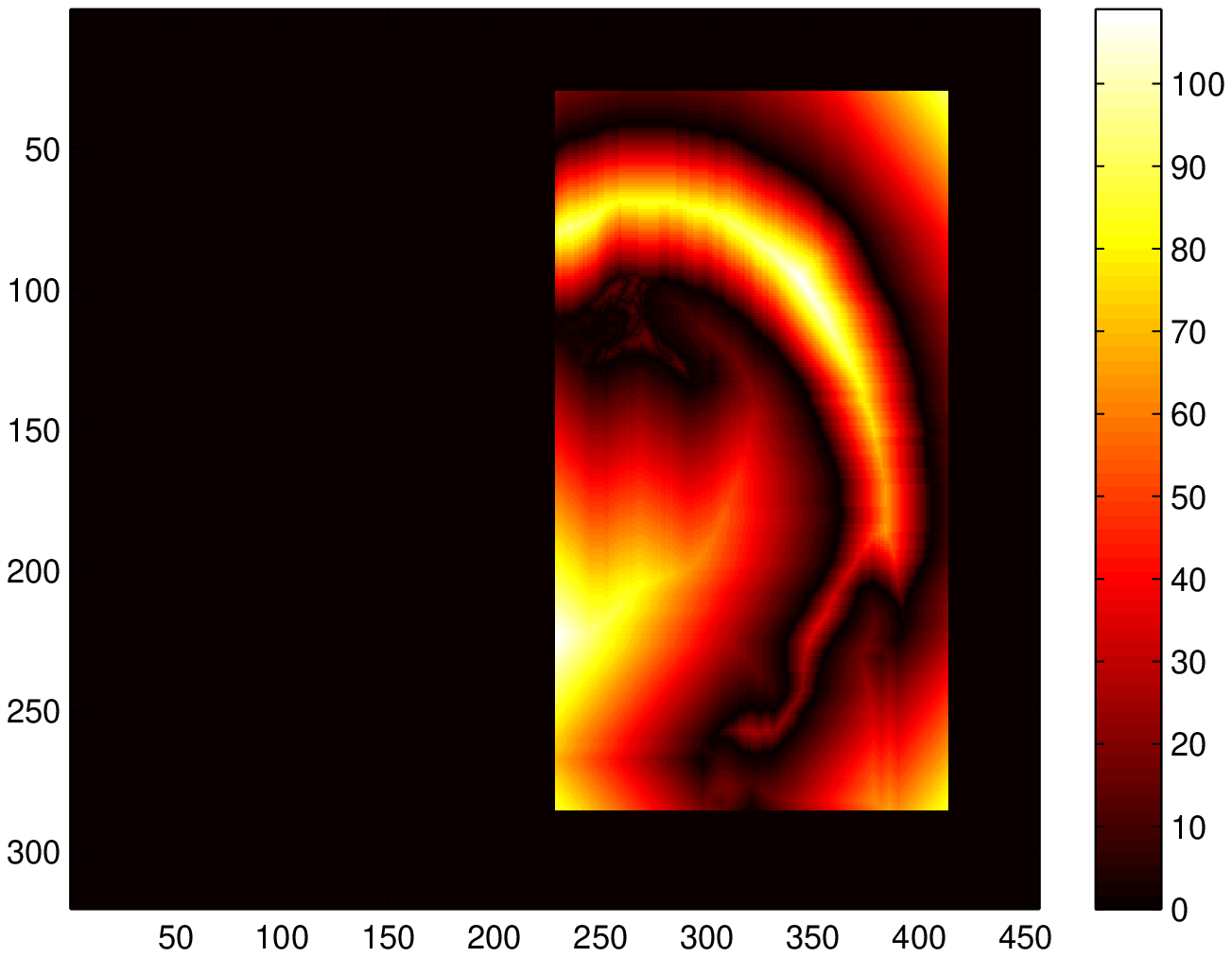}
\caption{A coronal section of the eikonal distances to the surface of the cerebral cortex. Distances in the 
colorbar are expressed in multiples of the voxel side, which is 25 microns.}
\label{fig:cortexEikonal}
%\end{minipage}
\end{figure}

%\begin{figure}
%\begin{minipage}{0.35\textwidth} 
Figure \ref{fig:cortexEikonal} is a coronal section of the eikonal
distance function to the boundary of the cerebral cortex in the mouse
brain. By applying a mask onto the interior of the region, one can
extract the surface inside the cerebral cortex that maximizes the
distance to the boundary on any coronal section. This surface serves
as an average layer, or {\emph{average surface}} of the cerebral
cortex.\\
%\end{minipage}
%\begin{minipage}{0.59\textwidth}
\begin{figure}
\begin{minipage}{0.49\textwidth} 
We would like to sample this surface in a homogeneous way. As it is
not too singular, a reasonable tiling of the surface is induced by
projecting a regular 2D hexagonal lattice (pictured in blue on the
figure \ref{fig:cortexTiling}) onto the surface. The resulting sites
are shown in magenta.\\ 

Given such a site, five injections are
regularly distributed along a line oriented by the maximum gradient of
the eikonal ditance across this site, so that the needle cut the brain along the
shortest possible length. There are some boundary effects as the
left-hemisphere has a vertical boundary on the right side that is not
physical and can be caught by the algorithm.\\
\end{minipage}
\begin{minipage}{0.49\textwidth}
\centering
\includegraphics[width=4in]{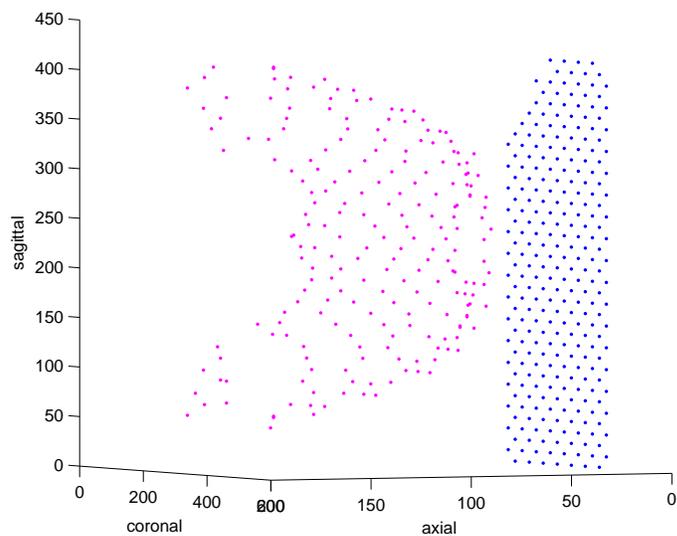}
\caption{The intersections of the needle trajectory and the average surface of the cerebral cortex, in magenta. They are obtained by projecting the two-dimensional grid shown in blue onto the average surface. }
\label{fig:cortexTiling}
\end{minipage}
\end{figure}

%\begin{figure}
%\begin{minipage}{0.4\textwidth} 
%A 3D plot of the injection targets, shown on  Figure (\ref{fig:dislocations}) looks locally like a crystal with many irregularities induces by boundaries of brain regions. On top of this crystal, the cortical injections look like ordered spines.\\
%\end{minipage}
%\begin{minipage}{0.58\textwidth} 

\newpage
\section*{Discussion}
\subsubsection*{Sampling constraint: pairwise distances within a region}
\begin{minipage}{0.4\textwidth} 
If more than one injection is made in a region, the minimum pairwise distance
between those injections should not be too different from $\sigma$, the 
spacing between vertices of a hexagonal packing that would fill the whole brain.
Discrepancies between these minimal distances and $\sigma$ come from the 
shape of the region, for example in cases were the region is much more elongated in 
one dimension than in the other two. The average value of those distances is 762 microns, and the standard deviation is 433 microns. 
\end{minipage}
\begin{minipage}{0.58\textwidth}
           %\includegraphics[width=\textwidth]{Satb2Coronal.eps} 
%\begin{figure}
%\centering
\includegraphics[width=4in]{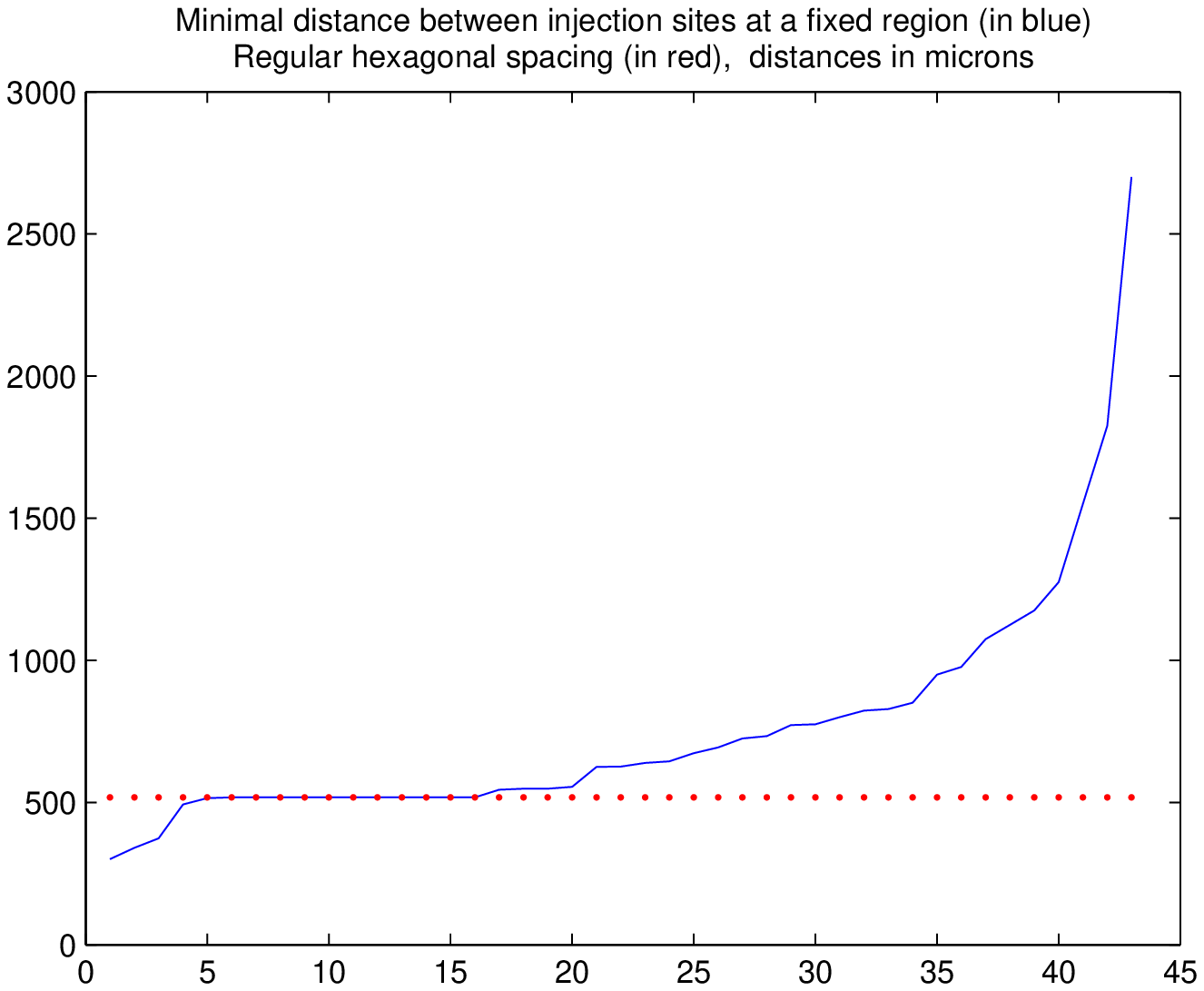}
%\caption{A plot of the minimal pairwise distances in regions with several injections
%(with the red line showing the value of $\sigma$.}
%\end{figure}
\end{minipage}

\subsubsection*{Separation constraints: distances to the boundaries}
\begin{minipage}{0.4\textwidth} 
As the security distance from the boundary has sometimes to be lowered
in order to fit a hexagonal packing with the desired number of 
vertices, and as the centroid of a region can happen to be close
to the boundary when the region is not convex, it is interesting to
compute the minimal distances between the injections made in a region,
and the boundary of this region (one just has to look up the values at
injection sites of the eikonal distance functions that were evaluated when the
security regions were determined). The mean distance to the boundary is 278 microns, close to $\sigma / 2$,
and the standard deviation is 72 microns. The exclusion of structures at the root of the tree, 
and the very procedure of excluding security regions from the computation, were expected to make 
the mean distance smaller than $\sigma$, with the same order of magnitude.\\
\end{minipage}
\begin{minipage}{0.58\textwidth}
           %\includegraphics[width=\textwidth]{Satb2Coronal.eps} 

%\begin{figure}
%\centering
\includegraphics[width=4in]{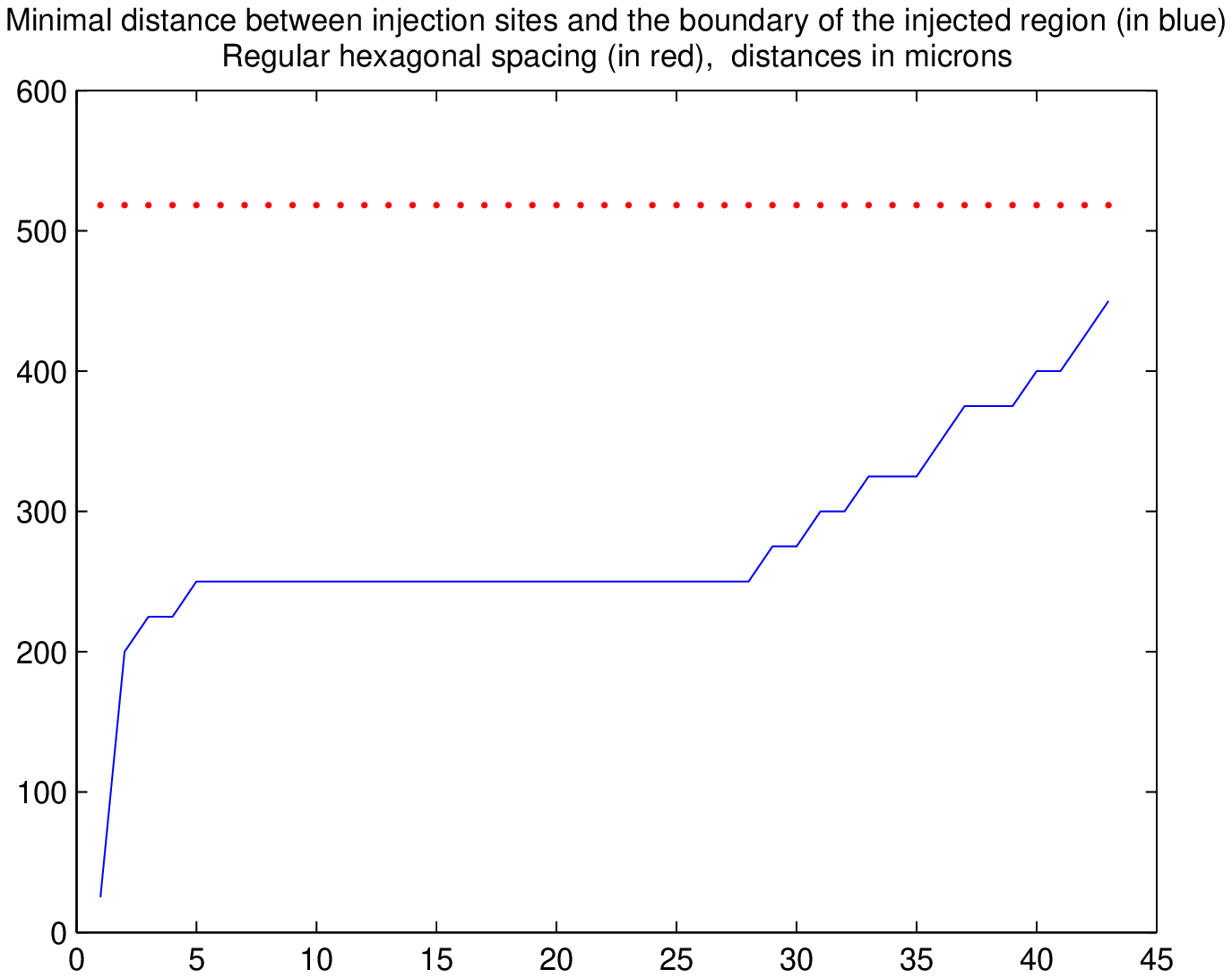}
%\caption{A plot of the minimal pairwise distances in regions with several injections
%(with the red line showing the value of $\sigma$.}
%\end{figure}
\end{minipage}\\
\subsubsection*{Distances between injections belonging to different regions}
\begin{minipage}{0.4\textwidth} 
For a given pair of regions indexed by $r$ and $s$ in the set of targeted regions,
 one can compute the three following quantities:\\
- the distance $d_{rs}$ between the averages of the voxels in the regions $r$ and $s$,\\
- the minimal distance $m_{rs}$ between an injection site in $r$, and an injection site in $s$,\\
- the maximal distances $M_{rs}$ between an injection site in $r$, and an injection site in $s$.\\
The plots of sorted values $m$ and $M$ should reproduce roughly the behaviour of $d$. This is indeed what
 we infer for the tubular shape of the three plots shown on the neighboring figure.\\
\end{minipage}
\begin{minipage}{0.58\textwidth}
%\begin{figure}
%\centering
\includegraphics[width=4in]{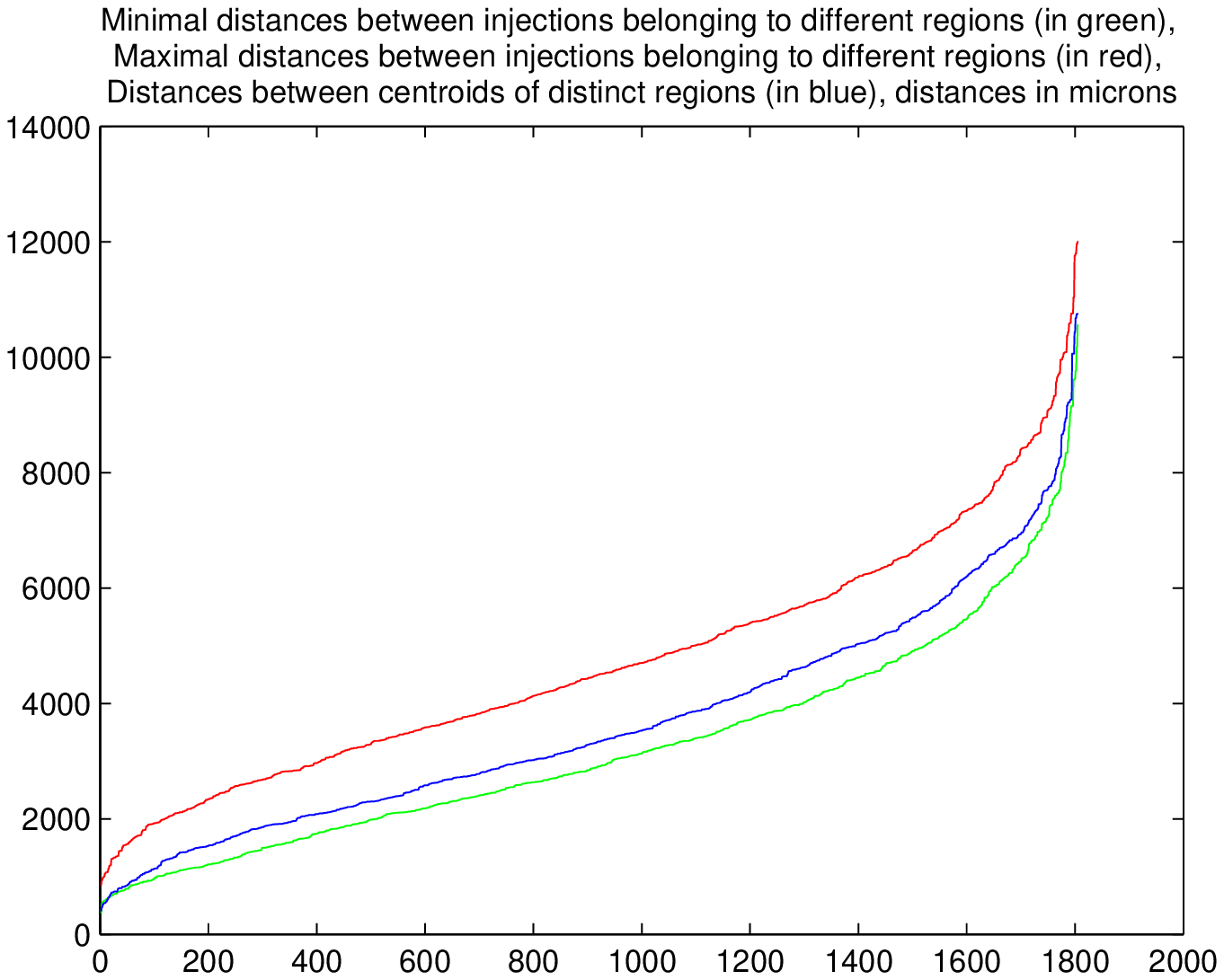}
%\caption{Minimal, average and maximal distances between sets injections grouped by targeted regions.}
%\end{figure}
\end{minipage}\\
\subsubsection*{Degrees of freedom and brain variability}
The placement of single injections at the centroid of the brain
regions (step 3.2 of the algorithm) would generalize naturally to larger sets of $k$ injections by
minimizing the sums of squared distances to the injections across the
region. This is the $k$-means algorithm \cite{kmeans}. Using this
algorithm, or any other clustering algorithm applied to the set of
coordinates of points within each brain region, would only change the
  step 3.2 of the algorithm. We did not choose to use $k$-means as the
initialization step of the algorithm would introduce some more
randomness.  Of course the coordinates that are returned by our
 code are not the single admissible solutions to the
algorithm. Each of the set of injections that consists of more than
one injection has three translation invariances within the region, one
for each coordinate axis. The amount of translation invariance
is equal to the distance by which the injections can be collectively moved without any of them touching
the boundary of the security region. Rotation degrees of freedom could 
also be considered, but this would break the alignment between 
planes containing several injections, and coronal sections of the atlas.\\ 

For practical purposes in the Mouse Brain Architecture project
\cite{MBA}, the precision of the computations has to be matched by the
surgical precision of the injections.  As the positions of boundaries
between regions are computed at a resolution of 25 microns, and as
security distances from those boundaries are typically a few hundreds
of microns, the geometric targets of the injections have to be reached
with a comparable precision. The alignment between the coordinate
systems in the Allen Atlas and in the live mouse is a first (geometric)
step to take in order to achieve such a precision.  A second (more
biological) step consists in an estimation of the animal-to animal
variability from a laser scan of the top of the skull.  Both steps
are taken algorithmically in a computer-guided stereotactic protocol
presented in \cite{stereotax}.\\

We treated the cerebral cortex seperately, and its characteristic
 layered structure will be sampled by columns of injections rather than by
pointwise injections. The cerebellar cortex also has a layered
structure, with layers separated by boundaries that could be described
in a useful way in terms of level sets of an eikonal
function. However, the boundary of the cerebellum has a folded profile
that could not be resolved on the grid we used. The many more
singularities of the boundary that are introduced by the layers would
also make the result of the algorithm less robust against brain
variability. The injection coordinates into cerebellar cortex were
therefore computed using the three-dimensional algorithm based on sphere packings.
 Addressing the folded structure of the cerebellum in a computational
way remains an open problem.

 \section*{Acknowledgments}
It is a pleasure to thank Kathleen Rockland for discussions. This
research is part of the Mouse Brain Architecture Project, supported by
grants 5RC1MJH088659, {\emph{The First Comprehensive Neural
    Connectivity Map of Mouse}}, and 5R01MH087988, {\emph{The Missing
    Circuit: The First Brainwide Connectivity Map for Mouse}}.

\end{document}